\documentclass[10pt]{iopart}
\usepackage{graphicx}
\usepackage{braket}
\begin{document}
\title[Microwave frequency conversion with a silicon nitride nanobeam oscillator]{Efficient microwave frequency conversion mediated by the vibrational motion of a silicon nitride nanobeam oscillator}
\author{J.~M.~Fink$^{1,2,3}$,
M.~Kalaee$^{1,2}$
R.~Norte$^{1,2}$\footnote{Present address: Department of Precision and Microsystems Engineering,
Delft University of Technology, Mekelweg 2, 2628CD Delft, The Netherlands},
A.~Pitanti$^{1,2}$\footnote{NEST, Istituto Nanoscienze-CNR and Scuola Normale Superiore, piazza San
Silvestro 12, 56127 Pisa (PI), Italy}
and
O.~Painter$^{1,2}$}
\address{$^1$ Kavli Nanoscience Institute and Thomas J. Watson, Sr., Laboratory of Applied Physics, California Institute of Technology, Pasadena, CA 91125, USA}
\address{$^2$ Institute for Quantum Information and Matter, California Institute of Technology, Pasadena, CA 91125, USA}
\address{$^3$ Institute of Science and Technology Austria (IST Austria), 3400 Klosterneuburg, Austria}
\ead{jfink@ist.ac.at}
\vspace{10pt}
\begin{indented}
\item[]Septempber 2018
\end{indented}

\begin{abstract}
Microelectromechanical systems and integrated photonics provide the basis for many reliable and compact circuit elements in modern communication systems. Electro-opto-mechanical devices are currently one of the leading approaches to realize ultra-sensitive, low-loss transducers for an emerging quantum information technology. 
Here we present an on-chip microwave frequency converter based on a planar aluminum on silicon nitride platform that is compatible with slot-mode coupled photonic crystal cavities. We show efficient frequency conversion between two propagating microwave modes mediated by the radiation pressure interaction with a metalized dielectric nanobeam oscillator. 
We achieve bidirectional coherent conversion with a total device efficiency of up to $\sim$~60\%, a dynamic range of $2\times10^9$ photons/s and an instantaneous bandwidth of up to 1.7~kHz. A high fidelity quantum state transfer would be possible if the drive dependent output noise of currently 
$\sim14$ photons$\ \cdot\ $s$^{-1}\ \cdot\ $Hz$^{-1}$ is further reduced. Such a silicon nitride based transducer is in-situ reconfigurable and could be used for on-chip classical and quantum signal routing and filtering, both for microwave and hybrid microwave-optical applications.
\end{abstract}
\vspace{2pc}
\ioptwocol
\maketitle

\section{Introduction}
Silicon nitride (Si$_3$N$_4$) thin films show exceptional optical and mechanical properties \cite{Zwickl2008}, and are
used in many microelectromechanical and photonic devices. The material's large bandgap \cite{Guo2018}, high power handling due to the absence of two-photon absorption in the telecom band \cite{Lacava2017} 
and the low absorption losses in Si$_3$N$_4$ thin films \cite{Barclay2006} make it an ideal candidate for many photonics applications, ranging from nonlinear optics \cite{Liu2013,Li2016}, to atom trapping \cite{Thompson2013,Yu2014b} and tests of quantum gravity \cite{Marshall2003}. The structural stability and high mechanical quality factor \cite{Southworth2009} of high tensile stress Si$_3$N$_4$ thin films grown by low-pressure chemical vapor deposition enables nanostructures to be patterned with extreme aspect ratios \cite{Cohen2013}
and form up to centimeter scale patterned membranes \cite{Duzzioni2005,Moura2018} with high reflectivity \cite{Norte2016,Chen2017}.
New soft clamping techniques make use of the tensile stress to maximize the mechanical quality factor \cite{Huang1998,Tsaturyan2017,Ghadimi2018}, allowing for an unprecedented regime in which quantum coherence can be reached for micromechanical systems even in a room temperature environment \cite{Sudhir2017}.
Additionally, slot mode 1D photonic crystal cavities have been developed using Si$_3$N$_4$ thin films to realize strong optomechanical interactions in a small mode volume and fully integrated on-chip \cite{Chan2009,Davanco2012b,Grutter2015,Norte2018}. 

In the microwave domain, Si$_3$N$_4$ is widely used for wiring capacitors and cross-overs. Early work focussed on the study of Si$_3$N$_4$ as a low loss dielectric to realize compact capacitive circuit elements operated in the quantum regime \cite{Paik2010}; however, the amorphous material and its surface are known to host two-level defects \cite{Faust2014}, such as hydrogen impurities with sizable dipole moments and life-times, which led to the observation of strong coupling between a single two-level system and a superconducting resonator \cite{Sarabi2015a,Fink2016,Sarabi2016b}. Nonetheless, due to its unique mechanical properties, high quality factor membranes \cite{Yuan2015,Noguchi2016} as well as micro-machined Si$_3$N$_4$ nanobeams \cite{Verbridge2008} have been coupled capacitively to superconducting resonators \cite{
Regal2008,
Rocheleau2010} in the context of cavity electromechanics. In the latter experiments the achievable coupling strength was fundamentally limited by the small participation ratio of the motional capacitance. We recently presented a new platform that uses the membrane itself as a low loss substrate for the microwave resonator, drastically lowering the parasitic circuit capacitance and maximizing the electromechanical coupling between a metalized silicon nitride nanobeam and a high impedance superconducting coil resonator. This allowed for high cooperativities and successfully demonstrating sideband cooling of the low MHz frequency nanobeam to the motional ground state \cite{Fink2016}. 

Silicon nitride membrane-based devices are currently the leading approach to couple optical and microwave systems \cite{Andrews2014,Higginbotham2018}. Realizing noiseless conversion with a mechanical oscillator \cite{Stannigel2010, Regal2011, Safavi-Naeini2011a, Barzanjeh2012, Clader2014, Tian2014, Zeuthen2017} would allow one to build transducers for quantum networks of superconducting processors connected via resilient and low loss optical fiber networks. Efficient wavelength conversion has been realized between optical wavelengths using silicon optomechanical crystals \cite{Hill2012}, between microwave frequencies using metallic drum resonators \cite{Andrews2015,Lecocq2016,Ockeloen-Korppi2016} and silicon nanobeams \cite{Barzanjeh2017}, and also between microwave and optical wavelength using silicon nitride membrane based Fabry-Perot cavities \cite{Andrews2014,Higginbotham2018}. Alternative approaches include the use of Josephson circuits for conversion in the microwave domain \cite{Lecocq2012,Sirois2015,Dmitriev2017}, Bragg scattering in silicon nitride rings \cite{Li2016} and dispersion engineering of silicon nitride waveguides \cite{Guo2018} in the optical domain. Coupling RF and microwave fields to optics has been achieved with membranes \cite{Bagci2014,Haghighi2018}, via a mechanical intermediary in combination with the piezoelectric effect and optomechanical interactions \cite{Bochmann2013, Balram2015, Forsch2018, Jiang2019},
and microwave to optics conversion has been proposed \cite{Tsang2010,Javerzac-Galy2016,Rueda2019b} and realized with high bandwidth via the electro-optic effect \cite{Rueda2016,Fan2018}. 

It is an outstanding challenge to realize an on-chip integrated microwave to optics converter based on the radiation pressure (optomechanical) interaction alone (i.e., on both the microwave and optical side of the converter). Mechanical systems offer the potential to fully separate the sensitive optical modes (superconductors cause optical loss) from the equally sensitive superconducting circuits (optical light generates quasi-particles in superconductors); for example using phononic waveguides. Here we present a coherent microwave frequency converter on the aluminum-on-Si$_3$N$_4$ platform \cite{Fink2016} that is compatible with on-chip optomechanics \cite{Davanco2012b,Grutter2015}. In the future this approach could be used to realize conversion between microwave and optical fields, or to implement low voltage modulation and fully electrical tunability in Si$_3$N$_4$-based photonic devices. 

\section{Implementation}
\subsection{Physics}
\begin{figure}[t]
\begin{center}
\includegraphics[width=1\columnwidth]{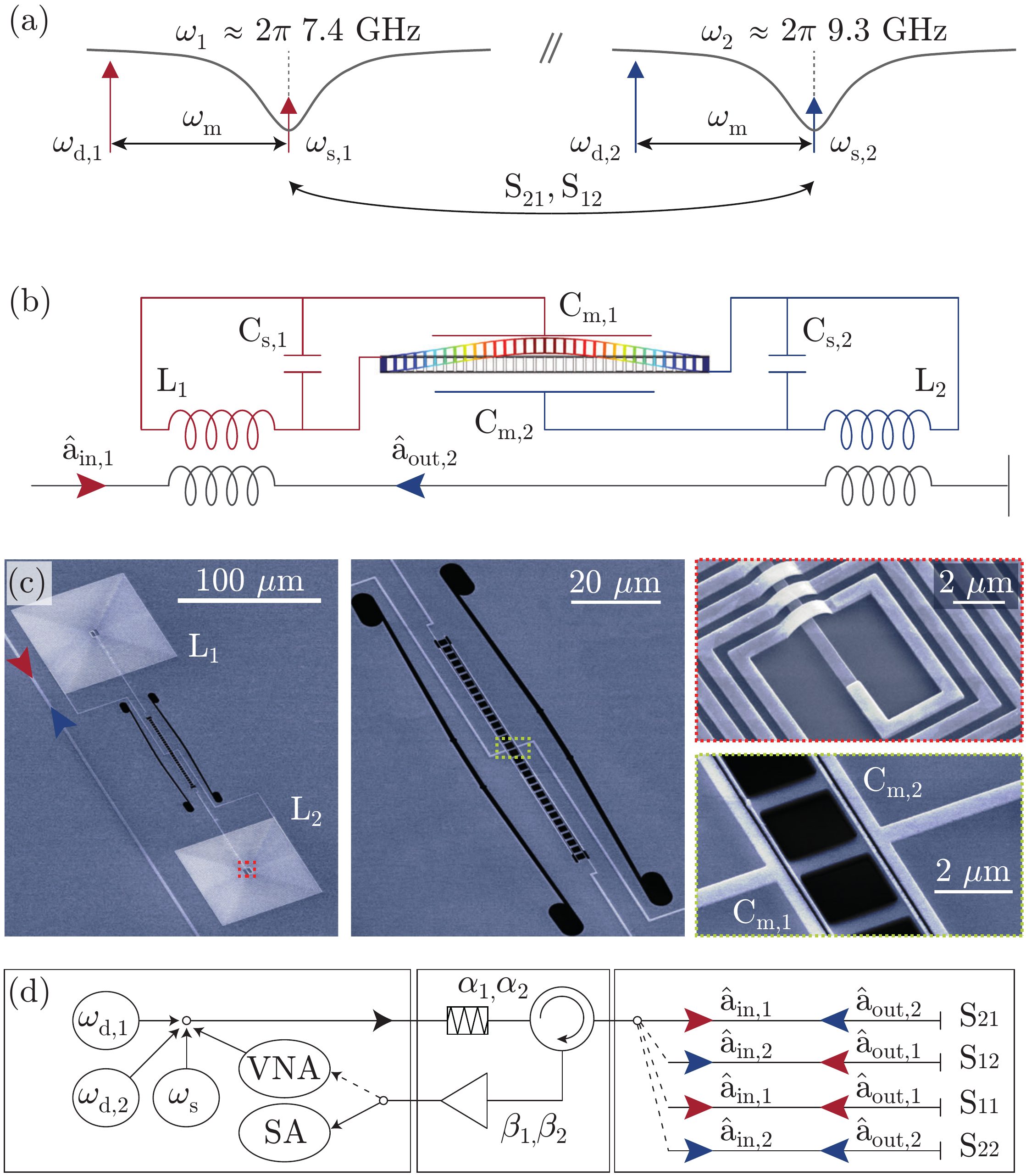}
\caption{
(a), Schematic presentation of the frequency conversion. The spectral density of the two microwave resonators at frequencies $\omega_i$ (black lines), the strong drive tones at frequencies $\omega_{\mathrm{d,}i}=\omega_i-\omega_\mathrm{m}$ (long red and blue arrows) and the signal tones at the optimal frequencies $\omega_{\mathrm{s,}i}=\omega_i$ (short red and blue arrows) for $i=\{1,2\}$, as well as the conversion scattering parameters $S_{21}$ and $S_{12}$ are indicated.   
(b), Circuit diagram of the converter. The silicon nitride nanobeam in-plane fundamental mode displacement (color indicates displacement amplitude) is coupled capacitively via its two modulated capacitances $C_{\mathrm{m,}i}$ to two parallel inductance-capacitance resonators realized with high characteristic impedance planar spiral inductors with the inductances $L_i$ and the stray capacitances $C_{\mathrm{s,}i}$. The two resonant circuits are coupled inductively to a transmission line to couple in and out the propagating microwave modes $\hat{a}_{\mathrm{in,}i}$ and $\hat{a}_{\mathrm{out,}i}$. 
(c), False color scanning electron micrograph of the converter device with thin-film aluminum (white) on suspended silicon nitride membrane (blue). Mechanical beam, cross-over and capacitor region are shown enlarged.
(d), Experimental setup. Three microwave sources and one vector network analyzer (VNA) output are combined at room temperature, attenuated by $\alpha_i$ and coupled to the device at about 12~mK using semirigid coaxial cables, a low loss printed circuit board and an on-chip coplanar waveguide. The reflected signals at the two frequencies of interest are routed to the output path using a cryogenic circulator and after passing another isolator (not shown) are amplified by $\beta_i$ at the 4 Kelvin stage and also at room temperature before detection with either a spectrum analyzer (SA) or the VNA input.
} \label{fig1}
\end{center} 
\end{figure}

We realize a system where one mechanical oscillator mode with frequency $\omega_\mathrm{m}$ and damping rate~$\gamma_\mathrm{m}$ is coupled to two electromagnetic resonator modes with resonance frequencies $\omega_i$ and linewidths $\kappa_i$  ($i=\{1,2\}$) via the optomechanical radiation pressure interaction as proposed in Refs.~\cite{Regal2011, Safavi-Naeini2011a, Barzanjeh2012}. In the presence of two red detuned classical drive fields $\alpha_{\mathrm{d,}i}$ near the red sideband of the respective microwave mode at $\omega_{\mathrm{d,}i}=\omega_{i}-\omega_\mathrm{m}$ the parametric interaction can be linearized and described by the sum of two beam splitter type interactions that allow to swap excitations between the mechanical and the two electromagnetic modes, see Fig.~\ref{fig1}(a). In the resolved-sideband limit~($\omega_\mathrm{m}\gg \kappa_{i},\gamma_\mathrm{m}$) the linearized electromechanical Hamiltonian in the rotating frames and the rotating wave approximation is given by
\begin{equation}\label{hamiltonian1}
H=\sum_{i=1,2}\hbar \Delta_i \hat{a}^{\dagger}_i \hat{a}_i+\hbar \omega_\mathrm{m} \hat{b}^{\dagger} \hat{b}+\sum_{i=1,2}\hbar g_i\Big(\hat{a}_i \hat{b}^{\dagger}+\hat{b} \hat{a}_i^{\dagger}\Big),
\end{equation}
where $\hat{a}_{i}$ is the annihilation operator for the microwave field mode, $\hat{b}$ is the annihilation operator of the mechanical mode, $\Delta_{i}=\omega_{i}-\omega_{\mathrm{d,}i}=\omega_\mathrm{m}$ is the detuning between the external driving field and the relevant resonator resonance, and $g_i=g_{{0},i}\sqrt{n_i}$ is the electromechanical coupling strength 
between the mechanical mode and resonator $i$ 
with $n_i=|\alpha_{\mathrm{d,}i}|^2=\frac{P_{\mathrm{in,}i}}{\hbar \omega_{\mathrm{d,}i}}\frac{4\kappa_{\mathrm{ex,}i}}{\kappa_i^2+4\Delta_i^2}$
the number of intra-resonator drive photons for the microwave input power with $P_{\mathrm{in,}i}$. 

The interaction terms of the Hamiltonian in Eq.~(\ref{hamiltonian1}) have two closely related effects. Optomechanical damping cools the mechanical motion with the rate $\Gamma_{i}=\frac{4g_i^2}{\kappa_i}$. At the same time this leads to the desired bidirectional photon conversion between two distinct electromagnetic frequencies. Using input-output theory, we can relate the itinerant input and output modes to the intra-cavity modes as $\hat{a}_{\mathrm{out,}i} = \sqrt{\kappa_{\mathrm{ex,}i}} \hat{a_i} - \hat{a}_{\mathrm{in,}i}$. In the photon conversion process, an input microwave signal at frequency~$\omega_{\mathrm{s,}1}$ with amplitude $\hat{a}_{\mathrm{in,}1}$
is down-converted to the mechanical mode at frequency $\omega_\mathrm{m}$, which corresponds to $\hat{b}^{\dagger} \hat{a}_1$ in Eq.~(\ref{hamiltonian1}). Next, during an up-conversion process the mechanical mode transfers its energy to the output of the other microwave resonator at frequency~$\omega_2$ and amplitude $\hat{a}_{\mathrm{out,}2}$,
which corresponds to $\hat{a}^{\dagger}_2 \hat{b}$ in Eq.~(\ref{hamiltonian1}). In this process the mechanical resonance is virtually populated, in the sense that the input signal is rapidly converted to the output signal, leaving little time for the population of the intermediate mechanics. Likewise, an input microwave signal at frequency~$\omega_2$ can be converted to frequency~$\omega_1$ by reversing the conversion process, see Fig.~\ref{fig1}(a). The Hermitian aspect of the Hamiltonian~(\ref{hamiltonian1}) makes this process bidirectional, without any unwanted loss, gain or noise. 

We define the photon conversion efficiency via the transmission scattering parameter, i.e.~as the ratio of the output-signal photon flux over the input-signal photon flux, $|S_{ij}|^2=\Big|\frac{\hat{a}_{\mathrm{out,}i}}{\hat{a}_{\mathrm{in,}j}}\Big|^2$. By solving the linearized Langevin equations we find that for signals on resonance with the microwave resonator, in the steady state 
the bidirectional conversion efficiency is given as \cite{Safavi2011}
\begin{equation}\label{transmission}
|S_{ij}|^2=|S_{ji}|^2=|T|^2=\eta_i\eta_j \frac{4 C_iC_j}{(1+C_i+C_j)^2},
\end{equation}
with $i,j=\{1,2\}$ the indices of the two modes. $C_{i}=\frac{\Gamma_{i}}{\gamma_\mathrm{m}}$ is the electromechanical cooperativity for resonator $i$
and $\eta_i=\frac{\kappa_{\mathrm{ex},i}}{\kappa_{i}}$ is the waveguide-to-resonator coupling ratio with $\kappa_i=\kappa_{\mathrm{in},i}+\kappa_{\mathrm{ex},i}$ the total damping rate, $\kappa_{\mathrm{ex},i}$ the decay rate into the waveguide and $\kappa_{\mathrm{in},i}$ the decay rate to any other mode. We also obtain a simple equation for the two reflection coefficients, which are given by
\begin{equation}\label{reflection}
|S_{ii}|^2=\Big(1-\frac{2\eta_i(1+C_j)}{1+C_i+C_j}\Big)^2.
\end{equation}
For lossless microwave cavities~$\eta_i=1$ and in the limit $C_1=C_2 \gg 1$ near unity photon conversion efficiency with $|T|^2=1$ and $|S_{11}|^2=|S_{22}|^2=0$ (zero reflection) can be achieved. The former condition ($C_1=C_2$) balances the photon-phonon conversion rates $\Gamma_{i}$, while the latter condition ($C_i \gg 1$) guarantees the mechanical damping rate is much smaller than the conversion rates $\gamma_\mathrm{m} \ll \Gamma_{i}$. In this limit the photon-to-photon conversion rate exceeds the mechanical damping rate - the rate at which phonons are exchanged with the noisy environment. This conversion process is coherent with the bandwidth given by $\Gamma=\gamma_\mathrm{m}+\Gamma_1+\Gamma_2$, which is the total back-action-damped linewidth of the mechanical resonator in the presence of the two microwave drive fields.

\subsection{Circuit}
We implement bidirectional frequency conversion in a circuit as shown in Fig.~\ref{fig1}(b). The two microwave resonators with resonance frequencies $\omega_1=7.444$~GHz and $\omega_2=9.308$~GHz are realized using two lumped element inductor-capacitor (LC) circuits formed from a planar spiral inductor of high impedance. The capacitance of these lumped element resonant circuits is defined by the sum of the stray capacitance of the circuit, which is dominated by the inductor stray capacitance, and the mechanically modulated capacitance. The two resonators are inductively coupled to a single physical port - a 50~$\Omega$ coplanar waveguide that is shorted to ground using a thin superconducting wire close the the two inductors.

\subsection{Device}
The described circuit is fabricated on the aluminum-on-Si$_3$N$_4$ platform similar to Ref.~\cite{Fink2016}. Here the entire aluminum circuit, which is shown in Fig.~\ref{fig1}(c), is suspended on a fully under-etched high-stress Si$_3$N$_4$ membrane on a high resistivity silicon chip. The inductors are realized as planar spiral inductors with a pitch of 1~$\mu$m which maximizes the obtained geometric inductance per unit length, and together with the small effective permittivity of the 60~nm thin membrane, minimizes the stray capacitance of the circuit. This in turn maximizes the obtained electromechnical couplings yielding measured values of $g_{0,1}/2\pi= 33$~Hz and $g_{0,2}/2\pi=44$~Hz for the fundamental in-plane mechanical mode of the patterned silicon nitride nanobeam with an intrinsic damping rate of $\gamma_\mathrm{m}/2\pi=7$~Hz at a resonance frequency of $\omega_\mathrm{m}/2\pi=4.118$~MHz. This is in good agreement with calculations based on perturbation theory and electromagnetic modeling of the 
electric field strength at the dielectric and metallic boundaries of the vacuum gap capacitor \cite{Pitanti2015}. While not measured in this work, the Si$_3$N$_4$ nanobeam has been designed as a phononic bandgap crystal that also localizes a high frequency acoustic defect mode \cite{Fink2016}. Very recently, quantum-level transduction of hypersonic mechanical motion could be demonstrated with a similar device \cite{Kalaee2019}.

\subsection{Setup}
The experiment is performed at 12~mK inside a dilution refrigerator. At room temperature we apply the drive and signal tones with low noise microwave sources and detect both reflection scattering parameters with a vector network analyzer (VNA) and the two transmission scattering parameters with a spectrum analyzer (SA), as shown in Fig.~\ref{fig1}(d). Inside the dilution refrigerator we distribute 50~$\Omega$ attenuators at various temperature stages to thermalize the electromagnetic mode temperature with the refrigerator temperature. We use one circulator to couple to the single physical port of the device in a reflection geometry. A second isolator is used to isolate the device from noise at 4 Kelvin where a commercial low noise HEMT amplifier is positioned. The four scattering parameters $S_{ii}$ between the two mode frequencies $\omega_i$ 
presented in this work refer to the ratios of the 4 propagating modes in a single on-chip waveguide as schematically shown in Fig.~\ref{fig1}(d).

\section{Characterization}
\subsection{Resonators}
As a first step we characterize the resonator properties using a VNA. 
The magnitudes of the measured complex reflection coefficients $S_{ii}$ are normalized with $\alpha_i \beta_i\rightarrow1$, which now corresponds to the scattering parameter at the position of the on-chip waveguide. Then the measured in-phase and quadrature phase components are fitted to the real and imaginary components of
\begin{equation}\label{resonator}
S_{ii}(\omega)=e^{-i (\phi+\omega \tau)} \left( 1 - \frac{\kappa_{\mathrm{ex,}i}}{\kappa_i/2+i(\omega_i-\omega)} \right),
\end{equation}
where $\phi$ is a global phase offset and $\tau\approx 50$~ns is the delay of the signal in our setup. The result for both resonator modes are shown in Fig.~\ref{fig2}(a). Here we plot the magnitude and phase of the measurement (blue points) and the fit (red lines) with excellent agreement. One can see that the resonators are both over-coupled, i.e. $\kappa_{\mathrm{ex,}i} > \kappa_{\mathrm{in,}i}$ as indicated by the full phase shift of $\sim 2 \pi$. The comparably very high quality factors of $Q_{\mathrm{in,}i}=\frac{\omega_{i}}{\kappa_{\mathrm{in,}i}}=\{2.2\times10^5, 5.5\times 10^4 \}$
enable the large waveguide coupling constants of up to $\eta_i=\{ 0.92, 0.68\}$ that are essential for the efficient conversion process (c.f.~Eq.~(\ref{transmission})). 
In general, we find  the intrinsic losses 
to be drive power dependent, likely due to saturation of two-level system absorption in the amorphous Si$_3$N$_4$~\cite{Paik2010}. For the  powers studied in this manuscript we determined coupling efficiencies in the range of $\eta_i=\{ 0.80-0.92, 0.54-0.68\}$.


\subsection{Two-mode EIT}
\begin{figure}[t]
\begin{center}
\includegraphics[width=1\columnwidth]{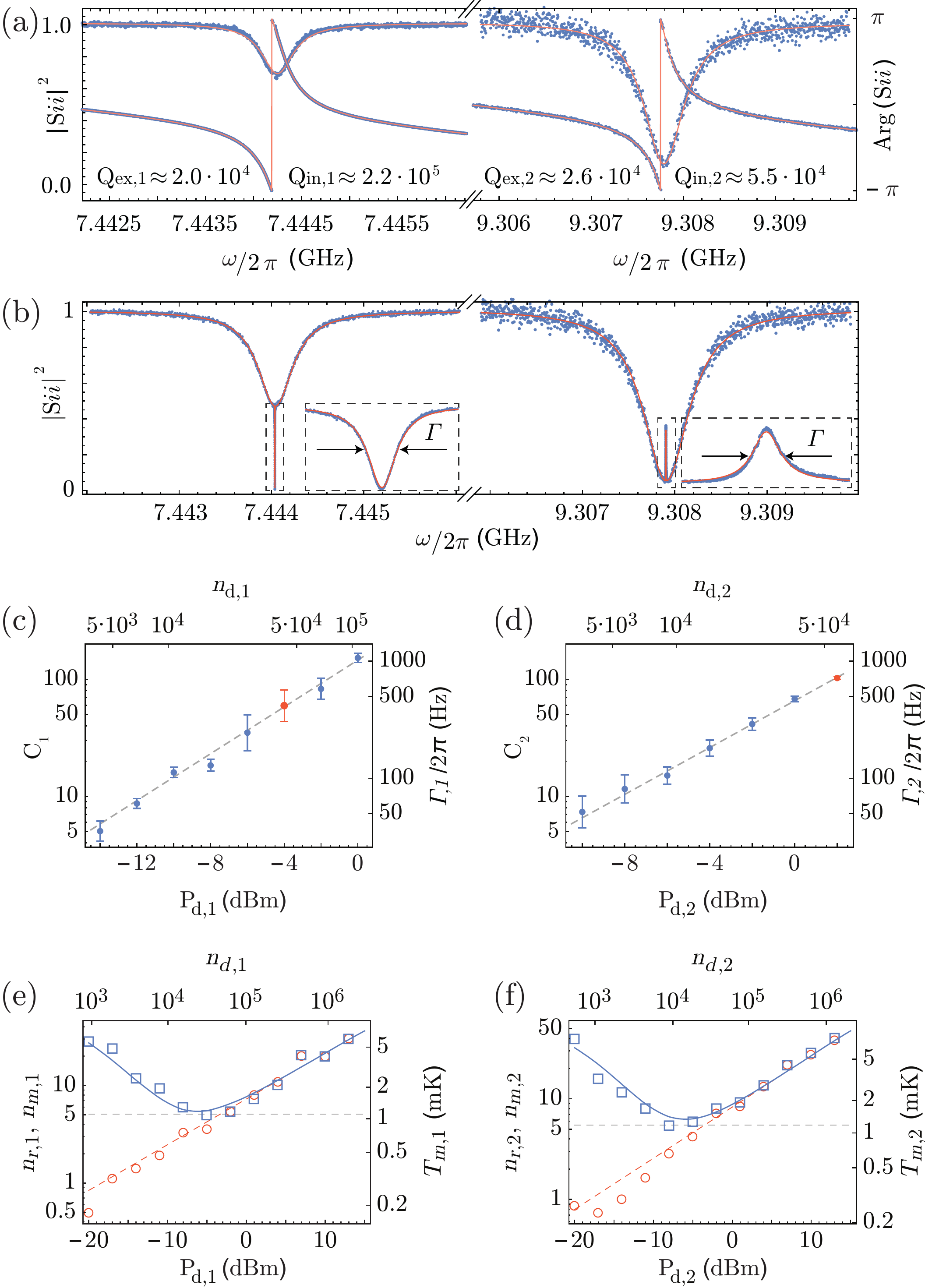}
\caption{
(a), Normalized reflected power (left axis) and phase (right axis) of a VNA measurement of the two microwave resonator modes (blue points) and a combined fit to the real and imaginary part of Eq.~(\ref{resonator}). The fitted extrinsic and intrinsic resonator quality factor are indicated. 
(b), Two-mode EIT measurement in the presence of two drive tones at $\omega_\mathrm{d,1}$ and $\omega_\mathrm{d,2}$ (blue points) and combined fit to the squared norm of Eq.~(\ref{omit}) for the two microwave modes. Insets show an enlarged view in the frequency direction (boxed area of main plot) of the mechanical response with identical total linewidth $\Gamma$. 
(c), and (d), Fitted optomechanical damping rate $\Gamma_i$ and cooperativity $C_i$ as a function of the drive power on resonator 1 (panel c) and 2 (panel d). Error bars show the standard deviation of the fitted cooperativity of one resonator inferred from multiple measurements with different drive powers applied to the other resonator. The data and fit shown in panel (b) was used for the data points indicated with red color.
Panels (e), and (f), show the mechanical $n_{\mathrm{m,}i}$ (blue squares) and resonator occupations $n_{\mathrm{r,}i}$ (red circles) as well as the mechanical noise temperatures $T_{\mathrm{m,}i}$ (right axis) as a function of red detuned drive power and the drive photon number. 
} \label{fig2}
\end{center} 
\end{figure}

As a second step we study the reflection scattering parameters measured with a weak probe tone using the VNA in the presence of two strong red-detuned drive tones. Here we observe a variant of optomechanically induced transparency \cite{Weis2010,Safavi-Naeini2011,Teufel2011}, an analog to electromagnetically induced transparency (EIT), where the mechanical sideband generated from the drives by the optomechanical interaction interfere with the weak probe tone from the VNA to modify the coherent resonator spectrum. In the case of two resonators and drive tones interacting with one mechanical mode we can model \cite{Hill2012a} and fit the measurements with 
\begin{equation}\label{omit}
S_{ii}(\omega) = 1- \frac{ \kappa_{\mathrm{ex,}i}\chi_{\mathrm{r,}i}\big(1+ g_j ^2 \chi_\mathrm{m} \chi_{\mathrm{r,}j} \big)}{1+ \chi_\mathrm{m} \sum_i g_i^2 \chi_{\mathrm{r,}i}}
\end{equation}
with $i,j=\{1,2\}$ the indices of the resonator modes, $\chi_{\mathrm{r,}i}^{-1}=\kappa_i/2 + i ( \Delta_i-\omega)$ the resonator susceptibilities and $\chi_\mathrm{m}^{-1}=\gamma_\mathrm{m}/2 + i (\omega_\mathrm{m}- \omega )$ the mechanical susceptibility. Figure~\ref{fig2}(b) shows a measurement (blue points) of the reflection scattering parameter in the vicinity of the two resonator modes together with a fit to Eq.~(\ref{omit}) (red lines). 

Similar to the case of standard EIT-type measurements, the formation of the peak or dip due to the mechanical mode with total linewidth $\Gamma$ (as indicated in the insets of Fig.~\ref{fig2}(b)) depends on the level of cooperativity $C$ and the degree of coupling $\eta$. However, in the present case the interpretation is more complicated because there are two terms that cause optomechanical damping. They affect each other and the double-EIT spectrum as a whole. For the chosen pump power and cooperativity combination, which is indicated in Fig.~\ref{fig2}(c) and (d) by red data points, we observe a full suppression of the reflected probe signal 
for resonator mode 1 Fig.~\ref{fig2}(b). This is a necessary condition for high efficiency conversion. In the case of resonator mode 2 on the other hand, which is only critically coupled to the waveguide ($\eta\sim0.5$), we observe a peak in the center of the resonator, indicating a finite reflection that results in a limited conversion efficiency. 

\subsection{Cooperativity}
We performed two-mode EIT measurements and analyses as presented above for all of the following pump power combinations: $P_{\mathrm{d,}1}$ from -14~dBm to 0~dBm and $P_{\mathrm{d,}2}$ from -10~dBm to 2~dBm, both in steps of 2~dB. Here the input power at the device $P_{\mathrm{in,}i}=P_{\mathrm{d,}i}-\alpha_i$ (on a log scale) is related to the shown drive powers via the attenuations $\alpha_i=\{69.0,70.4\}$~dB for the two mode frequencies of interest. For each power combination we fit both spectra to a single set of parameters (specifically $g_i$) and summarize the results in Fig.~\ref{fig2}(c) and (d). Shown are the mean and the statistical error of the cooperativities $C_i=\frac{\Gamma_i}{\gamma_\mathrm{m}}$. Using $\gamma_\mathrm{m}/2\pi=7$~Hz obtained from cooling measurements discussed below, we also back out the optomechanical damping $\Gamma_i=\frac{4 g_i^2}{\kappa_i}$ for each of the drive powers $P_{\mathrm{d,}i}$, and the intra-resonator drive photon numbers $n_{\mathrm{d,}i}$. We find that the dependence on drive power follows the expected behavior (dashed lines) based on the applied drive powers, attenuations and the calibrated $g_{0,i}$. The small error bars confirm that $C_1$ is not significantly affected by changing $P_{\mathrm{d,}2}$ and vice versa. The maximum $C_{1,2}\sim10^2$ that were obtained suggest that internal conversion efficiencies $\frac{|S_{ij}|^2}{\eta_1 \eta_2}\sim1$ can be achieved.

\subsection{Sideband cooling}
To estimate the noise generated by this device we perform motional sideband cooling \cite{Marquardt2007, Schliesser2008,Chan2011x,Teufel2011b,Fink2016} using each resonator mode independently, i.e.~with only one cooling tone at a time. The data is obtained by measuring the electronic noise spectrum due to the thermal Brownian motion of the nanobeam at the resonance frequency $\omega_\mathrm{m}$ with the linewidth $\Gamma=\Gamma_i+\gamma_\mathrm{m}$ as a function of the cooling drive tone power $P_{\mathrm{d,}i}$ applied at the optimal detuning $\Delta_i=\omega_i-\omega_\mathrm{m}$. The data shown in Fig.~\ref{fig2}(e) and f for modes 1 and 2 has been calibrated and analyzed as described in Ref.~\cite{Fink2016}. For small drive photon numbers we see a decrease of the phonon occupancies $n_{\mathrm{m,}i}$ (blue squares) in line with the expectations (blue lines), i.e. $n_{\mathrm{m,}i}=\frac{n_\mathrm{b}}{C_i+1}$ with $n_\mathrm{b}=60$ the mechanical bath corresponding to the refrigerator temperature of $\sim12$~mK and the intrinsic mechanical linewidth $\gamma_\mathrm{m}$ (blue squares). However, we also observe a power dependent increase of the noise floor with the bandwidth of the resonator mode \cite{Rocheleau2010}. Using a theoretical model \cite{Dobrindt2008, Fink2016} we fit the resonator occupations $n_{\mathrm{r,}i}$, which we show (red circles) together with a power law fit (dashed red line) in Fig.~\ref{fig2}(e) and (f). The power dependent resonator noise limits the minimum phonon occupation to $n_\mathrm{m}\sim5$ at drive powers of about -5~dBm, independent of which resonator mode is used for cooling. 

The residual thermal population of the mechanical oscillator ($n_\mathrm{m}\sim5$) and the two microwave resonators ($n_{\mathrm{r,}i}\sim4$) leads to incoherent added noise when the device is used as a transducer. 
In the limit $C_1\sim C_2 \gg 1$ and the realistic assumption that the waveguide modes are well thermalized and unpopulated, the noise added to any converted signal at the output of resonator port $i$ is given as \cite{Lecocq2016}
\begin{equation}\label{noise}
n_{\mathrm{add,}i}=\eta_i (n_{\mathrm{r,}1}+ n_{\mathrm{r,}2}+2 n_{\mathrm{m}}),
\end{equation}
which results in $n_{\mathrm{add,}i}\sim\{16, 12\}$ photons$\ \cdot\ $s$^{-1}\ \cdot\ $Hz$^{-1}$. For this estimate we have cautiously assumed the same mechanical population as measured with only a single drive tone, due to the cooling limitations imposed by the drive dependent resonator noise.
In addition, finite sideband resolution could lead to gain and amplified vacuum noise. With sideband resolution factors of $\frac{\kappa_i}{4\omega_\mathrm{m}}\leq10^{-3}$ this is expected to be negligible in the presented devices. 










\section{Wavelength conversion}

\subsection{Scattering parameters}
\begin{figure}[t]
\begin{center}
\includegraphics[width=1\columnwidth]{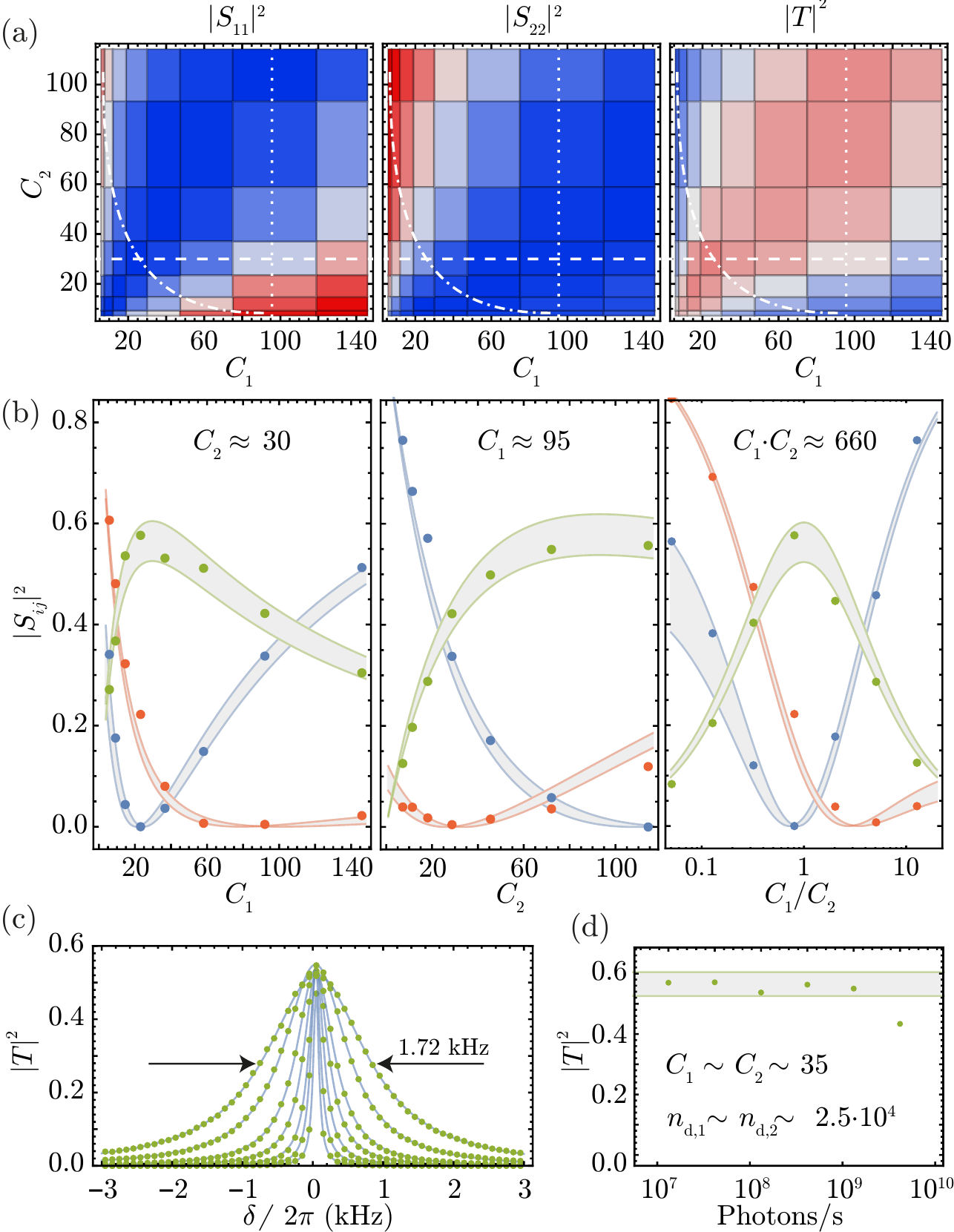}
\caption{
(a), Density plots of the two reflection and the bidirectional transmission scattering parameters (blue is low, red is high) as a function of both cooperativities $C_1$ and $C_2$. 
(b), Measured transmission (green dots), reflection from resonator 1 (blue dots) and resonator 2 (red dots) for the cooperativity values indicated by white lines (dashed, dotted, dashed-dotted) in panel (a). The bounds of the error bands shown (lines) are obtained from evaluating Eq.~\ref{transmission} and Eq.~\ref{reflection} for the minimum and maximum measured waveguide coupling efficiencies of $\eta_1={0.85,0.92}$ and $\eta_2={0.64,0.68}$. 
(c), The measured frequency dependence of the transmission (green dots) is shown as a function of the signal detuning $\delta$ together with a Lorentzian fit (lines) for different matching cooperativities. As expected, the bandwidth $\Gamma/2\pi$ of the conversion increases with $C=C_1=C_2$ and reaches a maximum of $1.72$~kHz. 
(d), Dynamic range of the converter. Measured transmission as a function of the converted signal power in units of photons/s. The shaded region indicates the expected conversion efficiency for the chosen cooperativity taking into account the uncertainty in the waveguide coupling efficiency.} 
\label{fig3}
\end{center} 
\end{figure}

In order to measure and quantify the efficiency of the wavelength converter we obtain the full set of scattering parameters for all 56 reported cooperativity combinations. The two reflection coefficients $|S_{ii}|^2$ are extracted from the center region of the two-mode EIT response shown in Fig.~\ref{fig2}(b). For the transmission we apply a coherent signal on resonance with one of the two microwave resonators and measure the mechanically transduced signal appearing in the center of the other resonator using a spectrum analyzer. Measuring this in both directions allows us to calibrate the product $|S_{21}| |S_{12}|=|T|^2$, i.e. the total bidirectional conversion efficiency \cite{Andrews2014}, using the product of both measured off-resonant reflection coefficients $\sqrt{\alpha_1 \beta_1} \sqrt{\alpha_2 \beta_2}\rightarrow1$, which were already used to normalize the reflection parameters in Fig.~\ref{fig2}(a) and (b). 
In practice, we sweep the signal frequency in a small range $\delta$ with a span on the order of the damped mechanical linewidth and extract the maximum value of the conversion efficiency. The result is shown in Fig.~\ref{fig3}(a) as a function of both cooperativities where red color indicates high and blue color indicates low values of the three scattering parameters. Qualitatively we see that matching the cooperativities leads to higher transmission and minimizes the reflection. 

In panel (b) of Fig.~\ref{fig3} we show the quantitative result for the three cooperativity combinations indicated with white lines in panel (a). In the first case we show all three $S$ parameters as a function of $C_1$ for $C_2\approx 30$. As expected we observe the maximum conversion efficiency of $|S_{ij}|^2$ and the minimum reflection $|S_{ii}|^2$ for $C_1\sim C_2$. In the second plot we fix a higher value ($C_1\approx95$) where we find that the optimal matching condition is relaxed, i.e.~the high conversion efficiency is achieved for a larger range of $C_2$. Finally, keeping the product constant $C_1\cdot C_2\approx660$ and plotting the scattering parameters as a function of the ratio $C_1/C_2$, the matching condition $C_1=C_2$ is very clear. Although the cooperativities are highest in the second case, the maximum conversion efficiency of $|S_{ij}|^2\approx 0.6$ is the same due to the predicted limitations imposed by the finite waveguide coupling efficiencies of $\eta_1\eta_2\approx0.63$. The estimated internal photon to photon conversion efficiency is $\frac{|S_{ij}|^2}{\eta_1 \eta_2}\approx 0.95$. These results are in excellent agreement with the simple theory presented in Eq.~(\ref{transmission}) and Eq.~(\ref{reflection}) shown as solid lines in Fig.~\ref{fig3}(b). 

\subsection{Bandwidth and dynamic range}
We have seen that the conversion efficiency saturates for $C_1=C_2\gg1$. The conversion bandwidth $\Gamma$ on the other hand is scaling with the cooperativity. Figure~\ref{fig3}(c) shows multiple signal frequency sweeps for different drive powers that match the cooperativity on both resonators. Here we achieve a maximum conversion bandwidth at the maximum efficiency of $\Gamma/2\pi=1.72$~kHz. 
The dynamic range is tested by extracting the maximum conversion efficiency as a function of the converted signal power for $C_1\approx C_2\approx 35$ and the result is shown in Figure~\ref{fig3}(d). We see no significant compression up to signal levels of $2\times10^9$ photons$/$s.


\section{Conclusion and Outlook}
In summary, we have presented a very versatile new dielectric nano-mechanical system that is suitable for efficient and near quantum limited wavelength conversion in the microwave frequency band. The unique properties of silicon nitride make it a natural platform for the realization of microwave-to-optical transducers and heralded entanglers \cite{Zhong2018}. Nevertheless, to achieve close to noise-free operation will require a deeper understanding of the electromagnetic and mechanical loss mechanisms, \cite{Yuan2015} and the thermal conductivity of low dimensional nano-beams \cite{Schwab2000} made from glassy materials like Si$_3$N$_4$ at temperatures in the range of only a few millikelvin \cite{Hauer2018}. This could lead to the observation of new and surprising physics along the way to fully operational quantum transducers and networks.



\footnotesize
\section*{Acknowledgments}
We thank Lukas Heinzle, Joe Redford, Marcelo Davan\c{c}o, Kartik Srinivasan and Greg McCabe for help in the early parts of this work and Shabir Barzanjeh for discussions. This research was supported by the DARPA MESO program, the ARO-MURI Quantum Opto-Mechanics with Atoms and Nanostructured Diamond (grant N00014-15-1-2761) and the Institute for Quantum Information and Matter, an NSF Physics Frontiers Center (grant PHY-1125565) with support of the Gordon and Betty Moore Foundation. A.P. was supported by a Marie Curie International Outgoing Fellowship within the 7th European Community Framework Program, NEMO (GA 298861). J.M.F acknowledges support from the European Research Council under grant agreement number 758053 (ERC StG QUNNECT), the Austrian Science Fund (FWF) through BeyondC (F71), a NOMIS foundation research grant, and the EU's Horizon 2020 research and innovation program under grant agreement number 732894 (FET Proactive HOT). 

\section*{References}
\bibliographystyle{naturemag}
\bibliography{FinkGroupBib_v6}

\begin{thebibliography}{10}
\expandafter\ifx\csname url\endcsname\relax
  \def\url#1{\texttt{#1}}\fi
\expandafter\ifx\csname urlprefix\endcsname\relax\def\urlprefix{URL }\fi
\providecommand{\bibinfo}[2]{#2}
\providecommand{\eprint}[2][]{\url{#2}}

\bibitem{Zwickl2008}
\bibinfo{author}{Zwickl, B.~M.} \emph{et~al.}
\newblock \bibinfo{title}{High quality mechanical and optical properties of
  commercial silicon nitride membranes}.
\newblock \emph{\bibinfo{journal}{Appl.\ Phys.\ Lett.}}
  \textbf{\bibinfo{volume}{92}}, \bibinfo{pages}{103125}
  (\bibinfo{year}{2008}).

\bibitem{Guo2018}
\bibinfo{author}{Guo, H.} \emph{et~al.}
\newblock \bibinfo{title}{Mid-infrared frequency comb via coherent dispersive
  wave generation in silicon nitride nanophotonic waveguides}.
\newblock \emph{\bibinfo{journal}{Nature Photonics}}
  \textbf{\bibinfo{volume}{12}}, \bibinfo{pages}{330--335}
  (\bibinfo{year}{2018}).

\bibitem{Lacava2017}
\bibinfo{author}{Lacava, C.} \emph{et~al.}
\newblock \bibinfo{title}{Si-rich silicon nitride for nonlinear signal
  processing applications}.
\newblock \emph{\bibinfo{journal}{Scientific Reports}}
  \textbf{\bibinfo{volume}{7}}, \bibinfo{pages}{22--} (\bibinfo{year}{2017}).

\bibitem{Barclay2006}
\bibinfo{author}{Barclay, P.~E.}, \bibinfo{author}{Srinivasan, K.},
  \bibinfo{author}{Painter, O.}, \bibinfo{author}{Lev, B.} \&
  \bibinfo{author}{Mabuchi, H.}
\newblock \bibinfo{title}{Integration of fiber-coupled high-q sinx microdisks
  with atom chips}.
\newblock \emph{\bibinfo{journal}{App. Phys. Lett.}}
  \textbf{\bibinfo{volume}{89}}, \bibinfo{pages}{131108}
  (\bibinfo{year}{2006}).

\bibitem{Liu2013}
\bibinfo{author}{Liu, Y.}, \bibinfo{author}{Davan\ifmmode~\mbox{\c{c}}\else
  \c{c}\fi{}o, M.}, \bibinfo{author}{Aksyuk, V.} \&
  \bibinfo{author}{Srinivasan, K.}
\newblock \bibinfo{title}{Electromagnetically induced transparency and wideband
  wavelength conversion in silicon nitride microdisk optomechanical
  resonators}.
\newblock \emph{\bibinfo{journal}{Phys. Rev. Lett.}}
  \textbf{\bibinfo{volume}{110}}, \bibinfo{pages}{223603}
  (\bibinfo{year}{2013}).

\bibitem{Li2016}
\bibinfo{author}{Li, Q.}, \bibinfo{author}{Davan\c{c}o, M.} \&
  \bibinfo{author}{Srinivasan, K.}
\newblock \bibinfo{title}{Efficient and low-noise single-photon-level frequency
  conversion interfaces using silicon nanophotonics}.
\newblock \emph{\bibinfo{journal}{Nature Photonics}}
  \textbf{\bibinfo{volume}{10}}, \bibinfo{pages}{406--} (\bibinfo{year}{2016}).

\bibitem{Thompson2013}
\bibinfo{author}{Thompson, J.~D.} \emph{et~al.}
\newblock \bibinfo{title}{Coupling a single trapped atom to a nanoscale optical
  cavity}.
\newblock \emph{\bibinfo{journal}{Science}} \textbf{\bibinfo{volume}{340}},
  \bibinfo{pages}{1202--1205} (\bibinfo{year}{2013}).

\bibitem{Yu2014b}
\bibinfo{author}{Yu, S.-P.} \emph{et~al.}
\newblock \bibinfo{title}{Nanowire photonic crystal waveguides for single-atom
  trapping and strong light-matter interactions}.
\newblock \emph{\bibinfo{journal}{App. Phys. Lett.}}
  \textbf{\bibinfo{volume}{104}}, \bibinfo{pages}{111103}
  (\bibinfo{year}{2014}).

\bibitem{Marshall2003}
\bibinfo{author}{Marshall, W.}, \bibinfo{author}{Simon, C.},
  \bibinfo{author}{Penrose, R.} \& \bibinfo{author}{Bouwmeester, D.}
\newblock \bibinfo{title}{{Towards Quantum Superpositions of a Mirror}}.
\newblock \emph{\bibinfo{journal}{Phys.\ Rev.\ Lett.}}
  \textbf{\bibinfo{volume}{91}}, \bibinfo{pages}{130401}
  (\bibinfo{year}{2003}).

\bibitem{Southworth2009}
\bibinfo{author}{Southworth, D.~R.} \emph{et~al.}
\newblock \bibinfo{title}{Stress and {S}ilicon {N}itride: {A} {C}rack in the
  {U}niversal {D}issipation of {G}lasses}.
\newblock \emph{\bibinfo{journal}{Phys.\ Rev.\ Lett.}}
  \textbf{\bibinfo{volume}{102}}, \bibinfo{pages}{225503}
  (\bibinfo{year}{2009}).

\bibitem{Cohen2013}
\bibinfo{author}{Cohen, J.~D.}, \bibinfo{author}{Meenehan, S.~M.} \&
  \bibinfo{author}{Painter, O.}
\newblock \bibinfo{title}{Optical coupling to nanoscale optomechanical cavities
  for near quantum-limited motion transduction}.
\newblock \emph{\bibinfo{journal}{Opt.\ Express}}
  \textbf{\bibinfo{volume}{21}}, \bibinfo{pages}{11227--11236}
  (\bibinfo{year}{2013}).

\bibitem{Duzzioni2005}
\bibinfo{author}{Duzzioni, E.~I.}, \bibinfo{author}{Villas-Boas, C.~J.},
  \bibinfo{author}{Mizrahi, S.~S.}, \bibinfo{author}{Moussa, M. H.~Y.} \&
  \bibinfo{author}{Serra, R.~M.}
\newblock \bibinfo{title}{Nonadiabatic geometric phase induced by a counterpart
  of the stark shift}.
\newblock \emph{\bibinfo{journal}{Europhys. Lett.}}
  \textbf{\bibinfo{volume}{72}}, \bibinfo{pages}{21--27}
  (\bibinfo{year}{2005}).

\bibitem{Moura2018}
\bibinfo{author}{Moura, J.~P.}, \bibinfo{author}{Norte, R.~A.},
  \bibinfo{author}{Guo, J.}, \bibinfo{author}{SchŠfermeier, C.} \&
  \bibinfo{author}{Gr\"oblacher, S.}
\newblock \bibinfo{title}{Centimeter-scale suspended photonic crystal mirrors}.
\newblock \emph{\bibinfo{journal}{Opt. Express}} \textbf{\bibinfo{volume}{26}},
  \bibinfo{pages}{1895--1909} (\bibinfo{year}{2018}).

\bibitem{Norte2016}
\bibinfo{author}{Norte, R.~A.}, \bibinfo{author}{Moura, J.~P.} \&
  \bibinfo{author}{Gr\"oblacher, S.}
\newblock \bibinfo{title}{Mechanical resonators for quantum optomechanics
  experiments at room temperature}.
\newblock \emph{\bibinfo{journal}{Phys. Rev. Lett.}}
  \textbf{\bibinfo{volume}{116}}, \bibinfo{pages}{147202}
  (\bibinfo{year}{2016}).

\bibitem{Chen2017}
\bibinfo{author}{Chen, X.} \emph{et~al.}
\newblock \bibinfo{title}{High-finesse fabry-perot cavities with bidimensional
  si3n4 photonic-crystal slabs}.
\newblock \emph{\bibinfo{journal}{Light: Science \&Amp; Applications}}
  \textbf{\bibinfo{volume}{6}}, \bibinfo{pages}{e16190--}
  (\bibinfo{year}{2017}).

\bibitem{Huang1998}
\bibinfo{author}{Huang, Y.~L.} \& \bibinfo{author}{Saulson, P.~R.}
\newblock \bibinfo{title}{Dissipation mechanisms in pendulums and their
  implications for gravitational wave interferometers}.
\newblock \emph{\bibinfo{journal}{Review of Scientific Instruments}}
  \textbf{\bibinfo{volume}{69}}, \bibinfo{pages}{544--553}
  (\bibinfo{year}{1998}).

\bibitem{Tsaturyan2017}
\bibinfo{author}{Tsaturyan, Y.}, \bibinfo{author}{Barg, A.},
  \bibinfo{author}{Polzik, E.~S.} \& \bibinfo{author}{Schliesser, A.}
\newblock \bibinfo{title}{Ultracoherent nanomechanical resonators via soft
  clamping and dissipation dilution}.
\newblock \emph{\bibinfo{journal}{Nature Nanotechnology}}
  \textbf{\bibinfo{volume}{12}}, \bibinfo{pages}{776--} (\bibinfo{year}{2017}).

\bibitem{Ghadimi2018}
\bibinfo{author}{Ghadimi, A.~H.} \emph{et~al.}
\newblock \bibinfo{title}{Elastic strain engineering for ultralow mechanical
  dissipation}.
\newblock \emph{\bibinfo{journal}{Science}} \textbf{\bibinfo{volume}{360}},
  \bibinfo{pages}{764--768} (\bibinfo{year}{2018}).

\bibitem{Sudhir2017}
\bibinfo{author}{Sudhir, V.} \emph{et~al.}
\newblock \bibinfo{title}{Quantum correlations of light from a room-temperature
  mechanical oscillator}.
\newblock \emph{\bibinfo{journal}{Phys. Rev. X}} \textbf{\bibinfo{volume}{7}},
  \bibinfo{pages}{031055} (\bibinfo{year}{2017}).

\bibitem{Chan2009}
\bibinfo{author}{Chan, J.}, \bibinfo{author}{Eichenfield, M.},
  \bibinfo{author}{Camacho, R.} \& \bibinfo{author}{Painter, O.}
\newblock \bibinfo{title}{Optical and mechanical design of a ``zipper''
  photonic crystal optomechanical cavity}.
\newblock \emph{\bibinfo{journal}{Opt. Express}} \textbf{\bibinfo{volume}{17}},
  \bibinfo{pages}{3802--3817} (\bibinfo{year}{2009}).

\bibitem{Davanco2012b}
\bibinfo{author}{Davan\c{c}o, M.}, \bibinfo{author}{Chan, J.},
  \bibinfo{author}{Safavi-Naeini, A.~H.}, \bibinfo{author}{Painter, O.} \&
  \bibinfo{author}{Srinivasan, K.}
\newblock \bibinfo{title}{Slot-mode-coupled optomechanical crystals}.
\newblock \emph{\bibinfo{journal}{Opt. Express}} \textbf{\bibinfo{volume}{20}},
  \bibinfo{pages}{24394--24410} (\bibinfo{year}{2012}).

\bibitem{Grutter2015}
\bibinfo{author}{Grutter, K.~E.}, \bibinfo{author}{Davan\c{c}o, M.~I.} \&
  \bibinfo{author}{Srinivasan, K.}
\newblock \bibinfo{title}{Slot-mode optomechanical crystals: a versatile
  platform for multimode optomechanics}.
\newblock \emph{\bibinfo{journal}{Optica}} \textbf{\bibinfo{volume}{2}},
  \bibinfo{pages}{994--1001} (\bibinfo{year}{2015}).

\bibitem{Norte2018}
\bibinfo{author}{Norte, R.~A.}, \bibinfo{author}{Forsch, M.},
  \bibinfo{author}{Wallucks, A.},
  \bibinfo{author}{Marinkovi\ifmmode~\acute{c}\else \'{c}\fi{}, I.} \&
  \bibinfo{author}{Gr\"oblacher, S.}
\newblock \bibinfo{title}{Platform for measurements of the casimir force
  between two superconductors}.
\newblock \emph{\bibinfo{journal}{Phys. Rev. Lett.}}
  \textbf{\bibinfo{volume}{121}}, \bibinfo{pages}{030405}
  (\bibinfo{year}{2018}).

\bibitem{Paik2010}
\bibinfo{author}{Paik, H.} \& \bibinfo{author}{Osborn, K.~D.}
\newblock \bibinfo{title}{Reducing quantum-regime dielectric loss of silicon
  nitride for superconducting quantum circuits}.
\newblock \emph{\bibinfo{journal}{Appl. Phys. Lett.}}
  \textbf{\bibinfo{volume}{96}}, \bibinfo{pages}{072505--}
  (\bibinfo{year}{2010}).

\bibitem{Faust2014}
\bibinfo{author}{Faust, T.}, \bibinfo{author}{Rieger, J.},
  \bibinfo{author}{Seitner, M.~J.}, \bibinfo{author}{Kotthaus, J.~P.} \&
  \bibinfo{author}{Weig, E.~M.}
\newblock \bibinfo{title}{Signatures of two-level defects in the
  temperature-dependent damping of nanomechanical silicon nitride resonators}.
\newblock \emph{\bibinfo{journal}{Phys. Rev. B}} \textbf{\bibinfo{volume}{89}},
  \bibinfo{pages}{100102} (\bibinfo{year}{2014}).

\bibitem{Sarabi2015a}
\bibinfo{author}{Sarabi, B.}, \bibinfo{author}{Ramanayaka, A.~N.},
  \bibinfo{author}{Burin, A.~L.}, \bibinfo{author}{Wellstood, F.~C.} \&
  \bibinfo{author}{Osborn, K.~D.}
\newblock \bibinfo{title}{Cavity quantum electrodynamics using a near-resonance
  two-level system: Emergence of the glauber state}.
\newblock \emph{\bibinfo{journal}{Applied Physics Letters}}
  \textbf{\bibinfo{volume}{106}} (\bibinfo{year}{2015}).

\bibitem{Fink2016}
\bibinfo{author}{Fink, J.~M.} \emph{et~al.}
\newblock \bibinfo{title}{Quantum electromechanics on silicon nitride
  nanomembranes}.
\newblock \emph{\bibinfo{journal}{Nature Communications}}
  \textbf{\bibinfo{volume}{7}}, \bibinfo{pages}{12396--}
  (\bibinfo{year}{2016}).

\bibitem{Sarabi2016b}
\bibinfo{author}{Sarabi, B.}, \bibinfo{author}{Ramanayaka, A.~N.},
  \bibinfo{author}{Burin, A.~L.}, \bibinfo{author}{Wellstood, F.~C.} \&
  \bibinfo{author}{Osborn, K.~D.}
\newblock \bibinfo{title}{Projected dipole moments of individual two-level
  defects extracted using circuit quantum electrodynamics}.
\newblock \emph{\bibinfo{journal}{Phys. Rev. Lett.}}
  \textbf{\bibinfo{volume}{116}}, \bibinfo{pages}{167002--}
  (\bibinfo{year}{2016}).
\newblock
  \urlprefix\url{https://link.aps.org/doi/10.1103/PhysRevLett.116.167002}.

\bibitem{Yuan2015}
\bibinfo{author}{Yuan, M.}, \bibinfo{author}{Cohen, M.~A.} \&
  \bibinfo{author}{Steele, G.~A.}
\newblock \bibinfo{title}{Silicon nitride membrane resonators at millikelvin
  temperatures with quality factors exceeding 108}.
\newblock \emph{\bibinfo{journal}{Appl. Phys. Lett.}}
  \textbf{\bibinfo{volume}{107}}, \bibinfo{pages}{263501--}
  (\bibinfo{year}{2015}).

\bibitem{Noguchi2016}
\bibinfo{author}{Noguchi, A.} \emph{et~al.}
\newblock \bibinfo{title}{Ground state cooling of a quantum electromechanical
  system with a silicon nitride membrane in a 3d loop-gap cavity}
  \textbf{\bibinfo{volume}{18}}, \bibinfo{pages}{103036--}
  (\bibinfo{year}{2016}).

\bibitem{Verbridge2008}
\bibinfo{author}{Verbridge, S.~S.}, \bibinfo{author}{Craighead, H.~G.} \&
  \bibinfo{author}{Parpia, J.~M.}
\newblock \bibinfo{title}{A megahertz nanomechanical resonator with room
  temperature quality factor over a million}.
\newblock \emph{\bibinfo{journal}{Appl. Phys. Lett.}}
  \textbf{\bibinfo{volume}{92}}, \bibinfo{pages}{013112--}
  (\bibinfo{year}{2008}).

\bibitem{Regal2008}
\bibinfo{author}{Regal, C.~A.}, \bibinfo{author}{Teufel, J.~D.} \&
  \bibinfo{author}{Lehnert, K.~W.}
\newblock \bibinfo{title}{Measuring nanomechanical motion with a microwave
  cavity interferometer}.
\newblock \emph{\bibinfo{journal}{Nat. Phys.}} \textbf{\bibinfo{volume}{4}},
  \bibinfo{pages}{555--560} (\bibinfo{year}{2008}).

\bibitem{Rocheleau2010}
\bibinfo{author}{Rocheleau, T.} \emph{et~al.}
\newblock \bibinfo{title}{Preparation and detection of a mechanical resonator
  near the ground state of motion}.
\newblock \emph{\bibinfo{journal}{Nature}} \textbf{\bibinfo{volume}{463}},
  \bibinfo{pages}{72--75} (\bibinfo{year}{2010}).

\bibitem{Andrews2014}
\bibinfo{author}{Andrews, R.~W.} \emph{et~al.}
\newblock \bibinfo{title}{Bidirectional and efficient conversion between
  microwave and optical light}.
\newblock \emph{\bibinfo{journal}{Nature Physics}}
  \textbf{\bibinfo{volume}{10}}, \bibinfo{pages}{321--326}
  (\bibinfo{year}{2014}).

\bibitem{Higginbotham2018}
\bibinfo{author}{Higginbotham, A.~P.} \emph{et~al.}
\newblock \bibinfo{title}{Harnessing electro-optic correlations in an efficient
  mechanical converter}.
\newblock \emph{\bibinfo{journal}{Nature Physics}} \bibinfo{pages}{--}
  (\bibinfo{year}{2018}).

\bibitem{Stannigel2010}
\bibinfo{author}{Stannigel, K.}, \bibinfo{author}{Rabl, P.},
  \bibinfo{author}{S\o{}rensen, A.~S.}, \bibinfo{author}{Zoller, P.} \&
  \bibinfo{author}{Lukin, M.~D.}
\newblock \bibinfo{title}{Optomechanical transducers for long-distance quantum
  communication}.
\newblock \emph{\bibinfo{journal}{Phys. Rev. Lett.}}
  \textbf{\bibinfo{volume}{105}}, \bibinfo{pages}{220501}
  (\bibinfo{year}{2010}).

\bibitem{Regal2011}
\bibinfo{author}{Regal, C.~A.} \& \bibinfo{author}{Lehnert, K.~W.}
\newblock \bibinfo{title}{From cavity electromechanics to cavity
  optomechanics}.
\newblock \emph{\bibinfo{journal}{Journal of Physics: Conference Series}}
  \textbf{\bibinfo{volume}{264}}, \bibinfo{pages}{012025}
  (\bibinfo{year}{2011}).

\bibitem{Safavi-Naeini2011a}
\bibinfo{author}{Safavi-Naeini, A.~H.} \& \bibinfo{author}{Painter, O.}
\newblock \bibinfo{title}{Proposal for an optomechanical traveling wave
  phonon-photon translator}.
\newblock \emph{\bibinfo{journal}{New J.\ Phys.}}
  \textbf{\bibinfo{volume}{13}}, \bibinfo{pages}{013017}
  (\bibinfo{year}{2011}).

\bibitem{Barzanjeh2012}
\bibinfo{author}{Barzanjeh, S.}, \bibinfo{author}{Abdi, M.},
  \bibinfo{author}{Milburn, G.~J.}, \bibinfo{author}{Tombesi, P.} \&
  \bibinfo{author}{Vitali, D.}
\newblock \bibinfo{title}{Reversible optical-to-microwave quantum interface}.
\newblock \emph{\bibinfo{journal}{Phys. Rev. Lett.}}
  \textbf{\bibinfo{volume}{109}}, \bibinfo{pages}{130503}
  (\bibinfo{year}{2012}).

\bibitem{Clader2014}
\bibinfo{author}{Clader, B.~D.}
\newblock \bibinfo{title}{Quantum networking of microwave photons using optical
  fibers}.
\newblock \emph{\bibinfo{journal}{Phys. Rev. A}} \textbf{\bibinfo{volume}{90}},
  \bibinfo{pages}{012324} (\bibinfo{year}{2014}).

\bibitem{Tian2014}
\bibinfo{author}{Tian, L.}
\newblock \bibinfo{title}{Optoelectromechanical transducer: Reversible
  conversion between microwave and optical photons}.
\newblock \emph{\bibinfo{journal}{ANNALEN DER PHYSIK}}
  \textbf{\bibinfo{volume}{527}}, \bibinfo{pages}{1--14}
  (\bibinfo{year}{2014}).

\bibitem{Zeuthen2017}
\bibinfo{author}{Zeuthen, E.}, \bibinfo{author}{Schliesser, A.},
  \bibinfo{author}{S{\o}rensen, A.~S.} \& \bibinfo{author}{Taylor, J.~M.}
\newblock \bibinfo{title}{Figures of merit for quantum transducers}.
\newblock \emph{\bibinfo{journal}{arXiv:1610.01099}}  (\bibinfo{year}{2017}).

\bibitem{Hill2012}
\bibinfo{author}{Hill, J.~T.}, \bibinfo{author}{Safavi-Naeini, A.~H.},
  \bibinfo{author}{Chan, J.} \& \bibinfo{author}{Painter, O.}
\newblock \bibinfo{title}{Coherent optical wavelength conversion via cavity
  optomechanics}.
\newblock \emph{\bibinfo{journal}{Nature Commun.}}
  \textbf{\bibinfo{volume}{3}}, \bibinfo{pages}{1196} (\bibinfo{year}{2012}).

\bibitem{Andrews2015}
\bibinfo{author}{Andrews, R.~W.}, \bibinfo{author}{Reed, A.~P.},
  \bibinfo{author}{Cicak, K.}, \bibinfo{author}{Teufel, J.~D.} \&
  \bibinfo{author}{W., L.~K.}
\newblock \bibinfo{title}{Quantum-enabled temporal and spectral mode conversion
  of microwave signals}.
\newblock \bibinfo{howpublished}{arXiv:1506.02296} (\bibinfo{year}{2015}).

\bibitem{Lecocq2016}
\bibinfo{author}{Lecocq, F.}, \bibinfo{author}{Clark, J.~B.},
  \bibinfo{author}{Simmonds, R.~W.}, \bibinfo{author}{Aumentado, J.} \&
  \bibinfo{author}{Teufel, J.~D.}
\newblock \bibinfo{title}{Mechanically mediated microwave frequency conversion
  in the quantum regime}.
\newblock \emph{\bibinfo{journal}{Phys. Rev. Lett.}}
  \textbf{\bibinfo{volume}{116}}, \bibinfo{pages}{043601}
  (\bibinfo{year}{2016}).

\bibitem{Ockeloen-Korppi2016}
\bibinfo{author}{Ockeloen-Korppi, C.~F.} \emph{et~al.}
\newblock \bibinfo{title}{Low-noise amplification and frequency conversion with
  a multiport microwave optomechanical device}.
\newblock \emph{\bibinfo{journal}{Phys. Rev. X}} \textbf{\bibinfo{volume}{6}},
  \bibinfo{pages}{041024} (\bibinfo{year}{2016}).

\bibitem{Barzanjeh2017}
\bibinfo{author}{Barzanjeh, S.} \emph{et~al.}
\newblock \bibinfo{title}{Mechanical on-chip microwave circulator}.
\newblock \emph{\bibinfo{journal}{Nature Communications}}
  \textbf{\bibinfo{volume}{8}}, \bibinfo{pages}{953--} (\bibinfo{year}{2017}).

\bibitem{Lecocq2012}
\bibinfo{author}{Lecocq, F.} \emph{et~al.}
\newblock \bibinfo{title}{Coherent frequency conversion in a superconducting
  artificial atom with two internal degrees of freedom}.
\newblock \emph{\bibinfo{journal}{Phys. Rev. Lett.}}
  \textbf{\bibinfo{volume}{108}}, \bibinfo{pages}{107001}
  (\bibinfo{year}{2012}).

\bibitem{Sirois2015}
\bibinfo{author}{Sirois, A.~J.} \emph{et~al.}
\newblock \bibinfo{title}{Coherent-state storage and retrieval between
  superconducting cavities using parametric frequency conversion}.
\newblock \emph{\bibinfo{journal}{Appl. Phys. Lett.}}
  \textbf{\bibinfo{volume}{106}}, \bibinfo{pages}{172603--}
  (\bibinfo{year}{2015}).

\bibitem{Dmitriev2017}
\bibinfo{author}{Dmitriev, A.~Y.}, \bibinfo{author}{Shaikhaidarov, R.},
  \bibinfo{author}{Antonov, V.~N.}, \bibinfo{author}{Hšnigl-Decrinis, T.} \&
  \bibinfo{author}{Astafiev, O.~V.}
\newblock \bibinfo{title}{Quantum wave mixing and visualisation of coherent and
  superposed photonic states in a waveguide}.
\newblock \emph{\bibinfo{journal}{Nature Communications}}
  \textbf{\bibinfo{volume}{8}}, \bibinfo{pages}{1352--} (\bibinfo{year}{2017}).

\bibitem{Bagci2014}
\bibinfo{author}{Bagci, T.} \emph{et~al.}
\newblock \bibinfo{title}{Optical detection of radio waves through a
  nanomechanical transducer}.
\newblock \emph{\bibinfo{journal}{Nature}} \textbf{\bibinfo{volume}{507}},
  \bibinfo{pages}{81--85} (\bibinfo{year}{2014}).

\bibitem{Haghighi2018}
\bibinfo{author}{Moaddel~Haghighi, I.}, \bibinfo{author}{Malossi, N.},
  \bibinfo{author}{Natali, R.}, \bibinfo{author}{Di~Giuseppe, G.} \&
  \bibinfo{author}{Vitali, D.}
\newblock \bibinfo{title}{Sensitivity-bandwidth limit in a multimode
  optoelectromechanical transducer}.
\newblock \emph{\bibinfo{journal}{Phys. Rev. Applied}}
  \textbf{\bibinfo{volume}{9}}, \bibinfo{pages}{034031} (\bibinfo{year}{2018}).

\bibitem{Bochmann2013}
\bibinfo{author}{Bochmann, J.}, \bibinfo{author}{Vainsencher, A.},
  \bibinfo{author}{Awschalom, D.~D.} \& \bibinfo{author}{Cleland, A.~N.}
\newblock \bibinfo{title}{Nanomechanical coupling between microwave and optical
  photons}.
\newblock \emph{\bibinfo{journal}{Nature Physics}}
  \textbf{\bibinfo{volume}{9}}, \bibinfo{pages}{712--716}
  (\bibinfo{year}{2013}).

\bibitem{Balram2015}
\bibinfo{author}{Balram, K.~C.}, \bibinfo{author}{Davan\c{c}o, M.},
  \bibinfo{author}{Song, J.~D.} \& \bibinfo{author}{Srinivasan, K.}
\newblock \bibinfo{title}{Coherent coupling between radio frequency, optical,
  and acoustic waves in piezo-optomechanical circuits}.
\newblock \bibinfo{howpublished}{arXiv:1508.01486} (\bibinfo{year}{2015}).
\newblock \bibinfo{note}{Balram2015}.

\bibitem{Forsch2018}
\bibinfo{author}{Forsch, M.} \emph{et~al.}
\newblock \bibinfo{title}{Microwave-to-optics conversion using a mechanical
  oscillator in its quantum groundstate}.
\newblock \emph{\bibinfo{journal}{arXiv:1812.07588}}  (\bibinfo{year}{2018}).

\bibitem{Jiang2019}
\bibinfo{author}{Jiang, W.} \emph{et~al.}
\newblock \bibinfo{title}{Efficient bidirectional piezo-optomechanical
  transduction between microwave and optical frequency}.
\newblock \emph{\bibinfo{journal}{arXiv:1909.04627}}  (\bibinfo{year}{2019}).

\bibitem{Tsang2010}
\bibinfo{author}{Tsang, M.}
\newblock \bibinfo{title}{Cavity quantum electro-optics}.
\newblock \emph{\bibinfo{journal}{Phys. Rev. A}} \textbf{\bibinfo{volume}{81}},
  \bibinfo{pages}{063837} (\bibinfo{year}{2010}).

\bibitem{Javerzac-Galy2016}
\bibinfo{author}{Javerzac-Galy, C.} \emph{et~al.}
\newblock \bibinfo{title}{On-chip microwave-to-optical quantum coherent
  converter based on a superconducting resonator coupled to an electro-optic
  microresonator}.
\newblock \emph{\bibinfo{journal}{Phys. Rev. A}} \textbf{\bibinfo{volume}{94}},
  \bibinfo{pages}{053815} (\bibinfo{year}{2016}).

\bibitem{Rueda2019b}
\bibinfo{author}{Rueda, A.}, \bibinfo{author}{Hease, W.},
  \bibinfo{author}{Barzanjeh, S.} \& \bibinfo{author}{Fink, J.~M.}
\newblock \bibinfo{title}{Electro-optic entanglement source for microwave to
  telecom quantum state transfer}.
\newblock \emph{\bibinfo{journal}{arXiv:1909.01470}}  (\bibinfo{year}{2019}).

\bibitem{Rueda2016}
\bibinfo{author}{Rueda, A.} \emph{et~al.}
\newblock \bibinfo{title}{Efficient microwave to optical photon conversion: an
  electro-optical realization}.
\newblock \emph{\bibinfo{journal}{Optica}} \textbf{\bibinfo{volume}{3}},
  \bibinfo{pages}{597--604} (\bibinfo{year}{2016}).

\bibitem{Fan2018}
\bibinfo{author}{Fan, L.} \emph{et~al.}
\newblock \bibinfo{title}{Superconducting cavity electro-optics: A platform for
  coherent photon conversion between superconducting and photonic circuits}.
\newblock \emph{\bibinfo{journal}{Science Advances}}
  \textbf{\bibinfo{volume}{4}} (\bibinfo{year}{2018}).

\bibitem{Safavi2011}
\bibinfo{author}{Safavi-Naeini, A.~H.} \& \bibinfo{author}{Painter, O.}
\newblock \bibinfo{title}{Proposal for an optomechanical traveling wave
  phonon–photon translator}.
\newblock \emph{\bibinfo{journal}{New Journal of Physics}}
  \textbf{\bibinfo{volume}{13}}, \bibinfo{pages}{013017}
  (\bibinfo{year}{2011}).

\bibitem{Pitanti2015}
\bibinfo{author}{Pitanti, A.} \emph{et~al.}
\newblock \bibinfo{title}{Strong opto-electro-mechanical coupling in a silicon
  photonic crystal cavity}.
\newblock \emph{\bibinfo{journal}{Opt. Express}} \textbf{\bibinfo{volume}{23}},
  \bibinfo{pages}{3196--3208} (\bibinfo{year}{2015}).

\bibitem{Kalaee2019}
\bibinfo{author}{Kalaee, M.} \emph{et~al.}
\newblock \bibinfo{title}{Quantum electromechanics of a hypersonic crystal}.
\newblock \emph{\bibinfo{journal}{Nature Nanotechnology}}
  \textbf{\bibinfo{volume}{14}}, \bibinfo{pages}{334--339}
  (\bibinfo{year}{2019}).

\bibitem{Weis2010}
\bibinfo{author}{Weis, S.} \emph{et~al.}
\newblock \bibinfo{title}{{Optomechanically Induced Transparency}}.
\newblock \emph{\bibinfo{journal}{Science}} \textbf{\bibinfo{volume}{330}},
  \bibinfo{pages}{1520--1523} (\bibinfo{year}{2010}).

\bibitem{Safavi-Naeini2011}
\bibinfo{author}{Safavi-Naeini, A.~H.} \emph{et~al.}
\newblock \bibinfo{title}{Electromagnetically induced transparency and slow
  light with optomechanics}.
\newblock \emph{\bibinfo{journal}{Nature}} \textbf{\bibinfo{volume}{472}},
  \bibinfo{pages}{69--73} (\bibinfo{year}{2011}).

\bibitem{Teufel2011}
\bibinfo{author}{Teufel, J.~D.} \emph{et~al.}
\newblock \bibinfo{title}{Circuit cavity electromechanics in the
  strong-coupling regime}.
\newblock \emph{\bibinfo{journal}{Nature}} \textbf{\bibinfo{volume}{471}},
  \bibinfo{pages}{204--208} (\bibinfo{year}{2011}).

\bibitem{Hill2012a}
\bibinfo{author}{Hill, J.~T.}, \bibinfo{author}{Safavi-Naeini, A.~H.},
  \bibinfo{author}{Chan, J.} \& \bibinfo{author}{Painter, O.}
\newblock \bibinfo{title}{Coherent optical wavelength conversion via cavity
  optomechanics}.
\newblock \emph{\bibinfo{journal}{Nat Commun}} \textbf{\bibinfo{volume}{3}},
  \bibinfo{pages}{1196--} (\bibinfo{year}{2012}).

\bibitem{Marquardt2007}
\bibinfo{author}{Marquardt, F.}, \bibinfo{author}{Chen, J.~P.},
  \bibinfo{author}{Clerk, A.~A.} \& \bibinfo{author}{Girvin, S.~M.}
\newblock \bibinfo{title}{Quantum theory of cavity-assisted sideband cooling of
  mechanical motion}.
\newblock \emph{\bibinfo{journal}{Phys.\ Rev.\ Lett.}}
  \textbf{\bibinfo{volume}{99}}, \bibinfo{pages}{093902}
  (\bibinfo{year}{2007}).

\bibitem{Schliesser2008}
\bibinfo{author}{Schliesser, A.}, \bibinfo{author}{Rivi\`{e}re, R.},
  \bibinfo{author}{Anetsberger, G.}, \bibinfo{author}{Arcizet, O.} \&
  \bibinfo{author}{Kippenberg, T.~J.}
\newblock \bibinfo{title}{Resolved-sideband cooling of a micromechanical
  oscillator}.
\newblock \emph{\bibinfo{journal}{Nature Phys.}} \textbf{\bibinfo{volume}{4}},
  \bibinfo{pages}{415--419} (\bibinfo{year}{2008}).

\bibitem{Chan2011x}
\bibinfo{author}{Chan, J.} \emph{et~al.}
\newblock \bibinfo{title}{Laser cooling of a nanomechanical oscillator into its
  quantum ground state}.
\newblock \emph{\bibinfo{journal}{Nature}} \textbf{\bibinfo{volume}{478}},
  \bibinfo{pages}{89--92} (\bibinfo{year}{2011}).

\bibitem{Teufel2011b}
\bibinfo{author}{Teufel, J.~D.} \emph{et~al.}
\newblock \bibinfo{title}{Sideband cooling of micromechanical motion to the
  quantum ground state}.
\newblock \emph{\bibinfo{journal}{Nature}} \textbf{\bibinfo{volume}{475}},
  \bibinfo{pages}{359--363} (\bibinfo{year}{2011}).

\bibitem{Dobrindt2008}
\bibinfo{author}{Dobrindt, J.~M.}, \bibinfo{author}{Wilson-Rae, I.} \&
  \bibinfo{author}{Kippenberg, T.~J.}
\newblock \bibinfo{title}{Parametric {N}ormal-{M}ode {S}plitting in {C}avity
  {O}ptomechanics}.
\newblock \emph{\bibinfo{journal}{Phys.\ Rev.\ Lett.}}
  \textbf{\bibinfo{volume}{101}}, \bibinfo{pages}{263602}
  (\bibinfo{year}{2008}).

\bibitem{Zhong2018}
\bibinfo{author}{Zhong, C.} \emph{et~al.}
\newblock \bibinfo{title}{Heralded generation and detection of entangled
  microwave--optical photon pairs}.
\newblock \emph{\bibinfo{journal}{arXiv:1901.08228}}  (\bibinfo{year}{2018}).

\bibitem{Schwab2000}
\bibinfo{author}{Schwab, K.}, \bibinfo{author}{Henriksen, E.~A.},
  \bibinfo{author}{Worlock, J.~M.} \& \bibinfo{author}{Roukes, M.~L.}
\newblock \bibinfo{title}{Measurement of the quantum of thermal conductance}.
\newblock \emph{\bibinfo{journal}{Nature}} \textbf{\bibinfo{volume}{404}},
  \bibinfo{pages}{974--977} (\bibinfo{year}{2000}).

\bibitem{Hauer2018}
\bibinfo{author}{Hauer, B.~D.}, \bibinfo{author}{Kim, P.~H.},
  \bibinfo{author}{Doolin, C.}, \bibinfo{author}{Souris, F.} \&
  \bibinfo{author}{Davis, J.~P.}
\newblock \bibinfo{title}{Two-level system damping in a quasi-one-dimensional
  optomechanical resonator}.
\newblock \emph{\bibinfo{journal}{Phys. Rev. B}} \textbf{\bibinfo{volume}{98}},
  \bibinfo{pages}{214303} (\bibinfo{year}{2018}).

\end{thebibliography}
\end{document}